\documentclass[letterpaper,conference]{IEEEtran}
\usepackage[T1]{fontenc}
\usepackage[latin9]{inputenc}
\usepackage{array}
\usepackage{float}
\usepackage{booktabs}
\usepackage{textcomp}
\usepackage{multirow}
\usepackage{amsmath}
\usepackage{amssymb}
\usepackage{graphicx}
\usepackage{setspace}

\makeatletter

\pdfpageheight\paperheight
\pdfpagewidth\paperwidth

\providecommand{\tabularnewline}{\\}
\newcommand{\lyxdot}{.}

\usepackage{colortbl}
\definecolor{lightgray}{gray}{0.8}

\usepackage{caption}
\usepackage{pifont}

\@ifundefined{showcaptionsetup}{}{%
 \PassOptionsToPackage{caption=false}{subfig}}
\usepackage{subfig}
\makeatother

\begin{document}

\title{Measurement and Prediction of Centrical/Peripheral Network Properties
based on Regression Analysis{\Large }\\
{\Large A Parametric Foundation for Performance Self-Management in
WSNs }}

\author{Adam Bachorek, Bagavathiannan Palanisamy and Jens B. Schmitt\smallskip{}
\\
{\normalsize disco | Distributed Computer Systems Lab}\\
{\normalsize University of Kaiserslautern, Germany}\\
{\normalsize \{bachorek, palanisamy, jschmitt\}@cs.uni-kl.de}}
\maketitle
\begin{abstract}
Predicting performance-related behavior of the underlying network
structure becomes more and more indispensable in terms of the aspired
application outcome quality. However, the reliable forecast of QoS
metrics like packet transfer delay in wireless network systems is
still a challenging task. Even though existing approaches are technically
capable of determining such network properties under certain assumptions,
they mostly abstract away from primal aspects that inherently have
an essential impact on temporal network performance dynamics. Also,
they usually require auxiliary resources to be implemented and deployed
along with the actual network components. In the course of developing
a lightweight measurement-based alternative for the self-inspection
and prediction of volatile performance characteristics in environments
of any kind, we selectively investigate the duration of message delivery
and packet loss rate against various parameters peculiar to common
radio network technologies like \emph{Wireless Sensor Networks (WSNs)}.
Our hands-on experiments reveal the relations between the oftentimes
underestimated medium access delay and a variety of main influencing
factors including packet size, backoff period, and number of neighbor
nodes contending for the communication medium. A closed formulation
of selected weighted drivers facilitates the average-case prediction
of inter-node packet transfer delays for arbitrary configurations
of given network parameters even on resource-scarce WSN devices. We
validate our prediction method against basic multi-hop networking
scenarios. Yield field test results proof the basic feasibility and
high precision of our approach to network property estimation in virtue
of self-governed local measurements and regression-based calculations
paving the way for a prospective self-management of network properties
based upon autonomous distributed coordination. \\
\end{abstract}

\section{Introduction}

Latest achievements in semi-conductor technology and \emph{micro-electro-mechanical
systems (MEMS)} have facilitated an increasing demand for low-cost
sensor devices that are miniaturized in shape but efficient in operation.
Further advances in the development of wireless communication techniques
and network equipment have made WSNs emerge as the seminal service
platform for realizing various valuable applications of today and
the ambitious vision of \emph{Ambient Intelligence (AI)} of tomorrow
\cite{2011-Buchmayr-surveysituationaware,2006-Benini-Wirelesssensornetworks:}.
At this stage, their small size and long life-span make WSN nodes
suitable for many application domains where dense sensing close to
physical phenomena as well as large-scale collection and exchange
of data is needed. Typically human-administered workaday applications
include biodiversity monitoring, emergency treatment support, facility
management, medical diagnostics and home automation \cite{2002-Akyildiz-WirelessSensorNetworks,2007-Akyildiz-surveywirelessmultimedia}.
In turn, the future design of AI envisions administration-free deployments
of even smaller, poly-functional and possibly also heterogeneous devices
gathering, processing and exchanging information from different sources
of the environment in order to exert influence on physical processes
in a further step. The involved ability of autonomous interaction
with ambient phenomena rather than with humans in the first place
is believed to be the crucial step towards pervasive environmental
control. However, in the long term, the implicated features are not
conceivable without a common notion of autonomy in regard to a self-governed
management of not only explicitly adjustable technical parameters
but also implicitly controllable functional properties as encountered
in actual hardware/software network deployments. At the same time,
the reliable and precise estimation of performance metrics without
human intervention constitutes a crucial feature of aspired applications
installed in, especially but not limited to, highly dynamic environments,
e.g., traffic control in over-crowded urban areas, rescue operations
in unstable disaster zones, let alone indoor surroundings with divers
mutually affecting home and facility instruments established for automation
purposes. From the point of view of various research endeavors, it
is obviously challenging to bring these stringent requirements in
line with present-day equipment that essentially requires manual superintendence
in order to fulfill any task at all due to lacking learning and adaptation
mechanisms \cite{2012-Movahedi-SurveyAutonomicNetwork,2011-KostasTsagkaris-Autonomicsinwireless}.
However, just as much as the concept of pervasive computing embodied
by ad hoc WSNs has enthused academia as well as the industry, the
notion of autonomic computing has witnessed great attention in regard
to the self-management of complex computer systems \cite{2003-Ganek-dawningautonomiccomputing,2005-Kephart-Researchchallengesautonomic}.
Autonomic networking that builds upon the same principles within the
context of the networking domain in order to tackle the increasing
complexity of ever-growing distributed (wireless) network systems,
has always played an important role to the same degree \cite{2009-Samaan-TowardsAutonomicNetwork,2007-BrendanJennings-Towardsautonomicmanagement}.
Whereas measurement-based estimation of metrics bears good prospects
in general \cite{2013-Volosencu-Efficiencyimprovementin,2005-Lim-Energyefficientself},
its consideration within the context of autonomic self-management
constitutes an even more promising approach. However, the estimation
of performance metrics in multi-hop networks like WSNs is not a straight-forward
task even for approved methodologies \cite{2007-Schmitt-ComprehensiveWorstCase,2011-RalfLuebben-foundationstochasticbandwidth}.
Furthermore, many works inherently rely on the inclusion of auxiliary
computing resources to tackle the implied issues of management due
to their common intricacy. However, we believe naturally autonomous
systems, as WSNs are aspired to become, shall only rely on their own
resources to fulfill their tasks all by themselves. Only this way
renders them useful for any kind of dynamic situations where additional
computing resources are just not available or integrable. That is
why we follow the approach to enable any given WSN to make use of
what it is being provided with, the enabling investigations of which
are the essential part our long-term research endeavors and to some
extent also part of this work. 

On this note, the remainder of this work is organized as follows.
In Section II we introduce our approach to the self-management of
performance properties including the needed terminology as the basis
for our investigations. Section III deals with details on our experimentation
methodology for the evaluation of network parameter impacts on major
network properties including a subsequent result analysis. As part
of Section IV, we discuss the proposed performance prediction technique,
review its general quality and also validate the results yield before
against a multi-hop scenario. Concluding remarks and prospects on
future works are eventually provided in Section V.

\begin{figure}
\centering{}\includegraphics[scale=0.4]{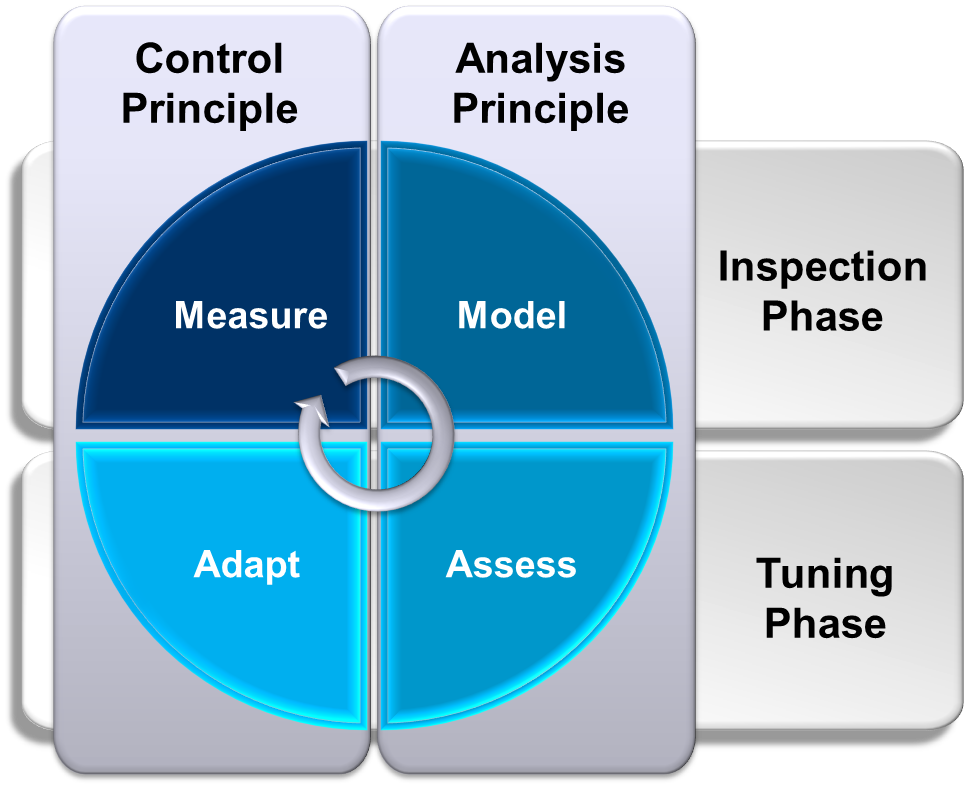}\caption{{\small \label{fig:Performance-Self-Management-Task-Cycle}Performance
Self-Management Task Cycle}}
\vspace{-0.6cm}
\end{figure}

\section{Performance Self-Management in WSNs - Parameters, Properties and
Metrics}

In general, the costs and benefits of autonomic networking not only
depend on the circumstances the corresponding hardware is subject
to but also on the correct handling of network (protocol) parameters
and the appropriate interpretation of encountered performance characteristics.
Since this inherently requires an efficient management of available
and controllable resources, it suggests itself to introduce a comprehensive
self-ability concept we recognize as \textbf{\emph{Performance Self-Management
(PSM)}}. 

In contrast, however also in complement, to other works that primarily
approach network autonomy by trying to establish overlay-like infrastructures
that are either attached to or even pervade the network of interest,
we take the approach of investigating principle methods that enable
node-local self-abilities, as we call them, from the node-local perspective
first. Nonetheless, we also share the view that during a transitional
period it will be required to provide the network entities with feedback
and functional support until all those methods fuse in a collaborative
manner rendering such networks operable on a fully autonomic basis.
That is why, we basically distinguish intrinsic from extrinsic PSM
in that we separate WSNs where most of the basic features are self-governed,
yet still require assistance from other entities for maintenance reasons,
e.g., human administrator during run-time operation, and WSNs that
keep-up the desired performance aspects on their own from the start
of the network application deployment until the end of its life-time.

In this spirit, we coin the term of \textbf{\emph{Intrinsic/Extrinsic
Performance Self-Management (I/EPSM)}} so as to define the scope of
our research directions within the context of wireless networked systems
like WSNs. To this end, we also identify a set of sub-routines that
need to be implemented on any I/EPSM-enabled node as part of that
concept. These modular self-abilities include measurement, modeling,
assessment and adaptation. The so-called \textbf{I/EPSM Task Cycle}
unites the four cornerstones of our approach the accomplishment of
which is based upon two fundamental principles that saturate two recurring
phases as illustrated in Figure \ref{fig:Performance-Self-Management-Task-Cycle}
and outlined in the following. 

\textbf{Inspection Phase: }The measurement task includes the inquiry
of relevant parameter values along with the capture of performance-related
statistics for the network properties of interest as indicated by
the requirements of the running application (\emph{Control Principle}).
The subsequent modeling task implies the identification of the best-fitting
model type for each of the network properties in question as well
as the instantiation of these models with the appropriate parameter
values as determined beforehand (\emph{Analysis Principle}).

\textbf{Tuning Phase:} The evaluation of the current performance quality
against given application requirements and, as the case may be, the
identification of those parameter constellations that need to and,
essentially, that can be changed so as to adhere to imposed network
property demands is part of the assessment task (\emph{Analysis Principle}).
At last, the agreed and optional adaption of the selected modifiable
parameters at the corresponding entities is accomplished in view of
the reinitialization and subsequent iteration of the overall mechanism
(\emph{Control Principle}). 

With regard to the frequency of the task cycle procedure, conceivable
solutions include time-based, entity-controlled and event-triggered
approaches that will depend on the network system application and
the aspired performance sustainment quality. Moreover, since the accomplishment
of at least a subset of all these modular tasks imposes a higher computational
as well as communicative burden on selected nodes involved in execution,
load-balancing mechanisms shall be applied for energy saving purposes
as part of future considerations. 

For the sake of transparent traceability of the upcoming details in
the subsequent sections, a clear terminology of the objects of investigation
is to be introduced next.

\subsection{Terminology}

In the context of I/EPSM, our terminology incorporates the general
distinction between two variable quantity types, network parameters
and network properties, both of which become manifest in metrics,
i.e., schemes of quantifiable measures for the characterization of
given physical phenomena. However, whereas\textbf{ }network\textbf{
parameters} include any accessible settings that need to be directly
or indirectly adjustable when acting as drivers for network property
control, network \textbf{properties}, on their part, involve any attributes
the values of which defy direct access and are obtainable merely by
means of measurements and/or computations based on statistical/analytical
inference in conjunction with concrete parameter values. 

We further differentiate the viewpoint on parameters and properties
from a centrical and a peripheral perspective. \textbf{Centrical parameters}
refer to intra-node/node-local variables the values of which can be
determined and tuned without any interaction with other entities (e.g.,
power amplifier level, packet size, backoff period, ...) while \textbf{peripheral
parameters} encompass inter-node/network-wide system variables usually
accessible and modifiable only by virtue of collaboration (e.g., number
of active nodes in cluster, network topology, number of neighbor connections,
...). Analogously, \textbf{centrical properties} pertain to measurable
or computable attributes that relate to the status of a single network
node (e.g., medium access delay, in-system backlog, traffic arrival
rate, ...) in contrast to \textbf{peripheral properties} which are
concerned with attributes going beyond a single network entity up
to network-wide conditions (e.g., end-to-end delay, cluster packet
loss rate, network backlog, ...). 

In some cases, otherwise regulable network parameters can be accessed
but are immutable due to, e.g., management policies or might also
need to be measured and somehow deduced from captured phenomena, respectively
(e.g., channel frequency, node position, distance to neighbors, ...).
However, as long as they remain uncontrollable to the involved network
entity/entities, they are not considered drivers for any properties.
Nonetheless, they may serve as invariable predictors for property
value determination in the same manner as other measurable or calculable
properties themselves can transitively do. Eventually, the causal
relationship between given network attributes shall be subject to
the inference of a suitable abstract intuition of real circumstances
bound within a calculable construct, henceforth referred to as \textbf{performance
model}.

As part of this work, we turn our attention to a representative selection
of one peripheral and two centrical parameters to investigate their
relation to two fundamental network properties as detailed in the
following.

\subsection{Network Properties under Investigation }

In the work at hand we focus on two of the most prominent and also
descriptive network properties capable of revealing the perception
of what is known to affect the timeliness and disposability of data. 

As per general convention the \textbf{\emph{End-to-End Delay (E2ED)}}
in packet-switched communication networks encompasses the entire time
duration between the start of the sending process at the origin of
the data of interest, commonly denoted as the source node, and the
end of reception at the ultimate target referred to as the destination
node \cite{2011-Peterson-Computernetworks:systems,2006-Kurose-ComputerNetworks:Top}.
In between, several delay components, traditionally known as the processing,
queuing, transmission, and propagation delay, can be identified dividing
the E2ED into several time segments that deserve a more detailed contemplation
in support of our forthcoming examinations.
\begin{itemize}
\item \emph{Processing delay}: time to check received, treat backlogged
and prepare to-send packet of interest primarily depending on amount
of data to be processed
\item \emph{Queuing delay:} time packet is blocked from reaching destined
communication port depending on current system/network load and data
buffer size in the first place
\item \emph{Transmission delay: }time required to deliver entire packet
onto communication medium particularly depending on size of data to
be sent
\item \emph{Propagation delay:} time packet needs to traverse distance towards
destination exclusively depending on signal propagation speed within
given communication medium
\end{itemize}
While propagation delay is actually negligible in short-distance \emph{Personal
Area Networks (PANs) }like WSNs due to its marginal and usually constant
impact on the overall time lag compared to the other delay components,
queuing delay might become the largest driver for E2ED in store-and-forward
networks that rely on data buffers. Yet, WSNs were originally not
meant to emulate the behavior of networks with dedicated devices,
such as highly buffered routers in IP-based nets, in order to solve
fundamental networking issues like forwarding. They were intended
make avail themselves of enabling concepts, e.g., data-centric collaboration,
deployment redundancy and in-network processing rather than relying
on single-node resources for such purposes \cite{2007-Karl-Protocolsandarchitectures}.
According to this, a typical WSN node is endued with a tiny transceiver
buffer of only one packet per link direction \cite{2007-TexasInstruments-ChipconCC2420-}.
Having said that, the illusive absence of queuing delay for small-buffered
devices, ought not to be underrated especially when it comes to the
packet loss and its implications on E2ED. In fact, a piece of data
that does not reach its destination due to, e.g., network congestion
might induce further delay by retransmissions as required by loss-intolerant
network applications, not to mention the auxiliary timeout latencies
introduced by the very usage of ACK-based reliable data transfer mechanisms
\cite{2003-Schiller-Mobilecommunications}. 

With regard to the other two delay components, they are often assumed
to be deterministically predictable solely by the knowledge of the
hardware details, e.g., transmission speed of the radio transceiver,
clock frequency of the microcontroller and the amount of data to be
handled. However, this does not hold true for any functionality that
is based on probabilistic procedures like the access arbitration in
random-based shared-medium schemes as defined, e.g., in the IEEE 802.15.4
\emph{Medium Access Control (MAC)} protocol for low-rate WPANs \cite{2003-IEEE-802.15.4-Standard}.
And, since the implicated delay for accessing the medium is usually
regarded as part of the queuing delay, it is neglected in turn when
abstracting away from queuing effects in buffer-scarce WSNs. Yet,
regardless of which of the above components it is actually attributed
to, it constitutes a remarkable influencing factor that necessitates
a thorough elaboration and further clear terminological separation.
\begin{itemize}
\item \textbf{\emph{Packet Processing Delay (PPD)}}: time for in-system
preparation and passing of packet data through any layer of communication-related
functionality from (when receiving) and towards (when sending) medium
access controlling system module
\item \textbf{\emph{Medium Access Delay (MAD)}}: time spent on carrier sensing
in conjunction with medium arbitration up to event of accessible idle
medium based on underlying medium access mechanism
\item \textbf{\emph{Packet Transmission Delay (PTD)}}: time required to
modulate entire packet onto communication medium up to the last bit
of data
\item \textbf{\emph{Packet Sending Delay (PSD)}}: time to accomplish entire
sending process, i.e., sum of PPD, MAD and PTD 
\end{itemize}
In a sense, our terminology constitutes a more detailed extraction
of the traditional division of delay components for practical reasons
in view of the upcoming experimentation without loss of generality.
That is to say, it allows to consider the otherwise peripheral notion
of E2ED also from a centrical view of a single network entity. Indeed,
a source node is prospectively supposed to predict the E2ED of a data
packet to a single-hop destination as the sum of its locally determinable
PSD and the PPD as encountered at the destination node. For the multi-hop
destination case, the source node further adds the sum of the corresponding
PPD and PSD for any other intermediate node the data packet needs
to traverse. After all, only node-local knowledge about the constitution
of its surroundings including inquired or deduced parameters and properties
shall suffice to enable attribute prediction.

The \textbf{\emph{Packet Loss Rate (PLR)}}, on the other hand, is
yet another well-established network property that offers valuable
clues to the current state of the peripheral as well as centrical
view on performance. In a way, it can give significant feedback to
network entities about the implications of current parameter settings
on other network properties that it correlates with. 

On this note, we define the PLR to be the number of packets that are
not received at the destination node in relation to the number of
packets issued at the sending source node. As a matter of fact, packet
loss is hard to detect in network systems that renounce reliable data
transfer mechanisms as commonly encountered in many broadcast-based
WSN application scenarios. However, since the event of loss might
be the consequence of a variety of reasons, e.g., system failure,
channel interference, duty cycling, that are implicitly interrelated
to that metric, other properties or parameters might be consulted
to deduce the PLR from the point of nodal view.

\subsection{Network Parameters of the Evaluation Platform}

Considering all major network parameters that are common to all wireless
network technologies similar to the exemplary platform we evaluate
in this work, we opt for one peripheral and two centrical parameters
as the objects of investigation due to their most promising influence
on the network properties mentioned above. 

First and foremost, the total number of nodes constitutes an intuitive
influencing factor for any kind of functionality, especially in case
of WSNs that define their basic principles upon their numerical redundancy.
However, since we are only interested in aspects from the centrical
view of a single node, for the time being, we draw our attention to
the neighbors in the vicinity of the node of interest that mutually
contend for medium access the quantity of which shall be denoted as
the \textbf{\emph{Number of Contenders (N$_{C}$)}} from now on.\textbf{\emph{ }}

Secondly, the length of the data packets, henceforth the \textbf{\emph{Packet
Size (P$_{S}$)}}, containing application-specific payload as well
as protocol-related management information that needs to be transferred
from the source to the destination node is taken into account.

At last, we identify a centrical parameter with highest impact on
the medium access behavior and, collaterally, also on the overall
inter-node transport process, which is referred to as the \textbf{\emph{Backoff
Period (B$_{P}$)}}.

\section{Experimentation on Self-Measurements - \protect \\
A Centrical/Node-local Approach}

As part of the control principle within the inspection-phase of the
I/EPSM task cycle, network nodes participating in network activity
are supposed to capture any parameter values that are considered important
with respect to their impact on certain network properties. To this
end, we are particularly interested in all the performance-related
information that can be obtained during regular network operation,
i.e., collected or derived when sending, receiving or processing data
in the context of orderly application execution (e.g., RSSI, medium
access delay, traffic input, missing ACKs, ...) without the need for
auxiliary energy consuming network communication. 

According to this centrical/node-local approach, the subsequent experimentation
deals with the basic feasibility of node-local measurements of network
properties and adjustments of network parameters implemented on real-world
WSN nodes leaning solely on out-of-the-box features as provided by
the associated software/hardware platform. In this regard, we explore
the dependency of PSD and PLR on three adjustable network parameters
including \emph{B$_{P}$}, \emph{P$_{S}$} and \emph{N$_{C}$} as
specified in the previous section. Besides, in order to accommodate
most WSN scenarios and also the situation of so-called ``event showers''
that are typical for many WSN applications, we designed our testbed
environment and evaluation procedure correspondingly as detailed below.

\subsection{Environmental Setup and Methodology}

The evaluation platform used throughout the experiments is the notorious
MICAz/TinyOS solution, one of the most commonly referenced WSN node
implementations featuring high versatility in plenty of adjustable
parameters and latest protocol mechanisms for the investigation of
WSN applications. MICAz motes feature an 8~MHz microprocessor, 4~kB
of RAM, 128~kB of code memory, and an IEEE 802.15.4-compliant transceiver
for radio communication of 128~Byte packets with 250~kbit/s of maximum
transmission rate \cite{2006-Crossbow-MoteProcessorRadio}. As the
standard operating system, TinyOS 2.1.2 based on the NesC programming
language is applied for running the motes \cite{2003-Gay-nesCLanguage-Holistic}. 

TinyOS implements a set of link-level primitives as specified by B-MAC
including optional use of link-level ACKs, a duty cycling mode, and
a basic CSMA mechanism without RTS/CTS that includes adjustable parameters
for MAC. In the context of PSD measurements, these parameters play
a fundamental role as delay-relevant factors. As per its carrier-sensing
mode, after a first waiting time (initial backoff) each packet transmission
attempt of a mote stipulates a prior \emph{Clear Channel Assessment
(CCA)} for channel state testing based on an averaged noise sampling
mechanism \cite{2004-Polastre-Versatilelowpower}. If the CCA encounters
a busy channel, the sending node backs off for a certain amount of
time (congestion backoff). This backoff delay $d_{B}$ is randomly
drawn from an interval spanned by means of two parameters, the so-called
minimum backoff $B{}_{min}$ and the backoff period $B{}_{p}$, as
given in Eq. \ref{eq:Backoff_Interval} below.

\begin{equation}
d_{B}=\left(r\, mod\left(z\cdot B{}_{p}\right)\right)+B{}_{min}\label{eq:Backoff_Interval}
\end{equation}

where $r$ is a 16~bit random value, $z=31$ is the initial backoff
factor on first transmission attempt and $z=7$ is the congestion
backoff factor utilized if the channel is sensed busy during any transmission
attempt. In the latter case, $d_{B}$ is drawn repeatedly until the
medium is sensed idle again the overall lead time of which is recognized
as MAD (see Section 2.2). On average, the smaller congestion backoff
factor gives transmission reattempts higher priority over initial
sending trials for fairness reasons. After $d_{B}$ slot times of
deferment, where $t_{slot}=2\cdot t_{symbol}=32\,\text{\textmu s}$,
the node is ready to send. By default $B{}_{min}=B{}_{per}=10$ with
$B{}_{min}$ regarded as platform-specific covering all hardware-related
in-system guard spaces and turnaround times and, thus, needs to remain
constant for proper transceiver operation. Furthermore, its constant
value has far less influence on MAD in contrast to $B{}_{per}$ as
initial experiments had shown. In terms of adjustable packet sizes,
TinyOS implements IEEE 802.15.4-compliant protocol headers and footers
accounting to 13~Byte by default which allows for additional payload
lengths of up to 115~Byte, all in all referred to as P$_{S}$ (see
Section 2.2) in the following \cite{2009-Levis-TinyOSprogramming}.

Due to possible activity of co-existing technologies, e.g., WLAN operating
within the same frequency band, node transceivers are set to a channel
providing for least collision potential \cite{2007-Alliance-ZigBeeandWireless}.
The power amplifier is set to the highest level for sufficient transmission
range and uninterrupted connectivity of involved motes that are randomly
arranged in a star-shaped topology with a radius of less than 1~m
allowing for line-of-sight communication. For highly reliable and
accurate delay measurements, an explicit TinyOS hardware timer abstraction
named \emph{Alarm} with \textmu{}s precision and 32~bit of width
is applied (as per timing precision evaluation of TinyOS timer abstraction
settings in \cite{2008-Bachorek-EnablingAuthenticTransmissions}).
Besides, link-level ACKs are deactivated avoiding retransmissions
of non-acknowledged frames for the sake of unambiguous measurement
interpretation. All relevant configuration settings are summarized
in Figure \ref{tab:General-and-experimental-configuration-settings}
(left).

\begin{figure}
\begin{centering}
{\scriptsize }%
\begin{minipage}[t]{1\columnwidth}%
\begin{center}
{\scriptsize \vspace{0cm}
}%
\begin{tabular}{cc}
\toprule 
\multicolumn{2}{c}{\textbf{\scriptsize General Node Configuration}}\tabularnewline
\midrule
\midrule 
\textbf{\scriptsize Parameter} & \textbf{\scriptsize Setting}\tabularnewline
\midrule 
{\scriptsize B$_{min}$} & {\scriptsize 10 (320~\textmu{}s)}\tabularnewline
\midrule 
{\scriptsize PAL} & {\scriptsize 31 (0~dBm)}\tabularnewline
\midrule 
{\scriptsize Channel} & {\scriptsize 26 (2.48~GHz)}\tabularnewline
\midrule 
{\scriptsize ACKs} & {\scriptsize 0 (disabled)}\tabularnewline
\midrule 
{\scriptsize Duty Cycle} & {\scriptsize 1 (always on)}\tabularnewline
\bottomrule
\end{tabular}{\scriptsize ~~~}%
\begin{tabular}{cc}
\toprule 
\multicolumn{2}{c}{\textbf{\scriptsize Experiment Run Configuration}}\tabularnewline
\midrule
\midrule 
\textbf{\scriptsize Parameter} & \textbf{\scriptsize Setting}\tabularnewline
\midrule 
{\scriptsize B$_{P}$} & {\scriptsize {[}1,20{]} by 1 }\tabularnewline
\midrule 
{\scriptsize P$_{S}$ {[}Byte{]}} & {\scriptsize {[}20,120{]} by 5}\tabularnewline
\midrule 
{\scriptsize N$_{C}$} & {\scriptsize \{1,2,4,8\}}\tabularnewline
\midrule 
{\scriptsize Samples {[}packets{]}} & {\scriptsize 1000}\tabularnewline
\midrule 
{\scriptsize Event rate {[}1/s{]}} & {\scriptsize 10}\tabularnewline
\bottomrule
\end{tabular}\vspace{0.2cm}

\par\end{center}%
\end{minipage}
\par\end{centering}{\scriptsize \par}

\caption{{\small \label{tab:General-and-experimental-configuration-settings}General
and experimental configuration settings}}
\vspace{-0.5cm}
\end{figure}

\begin{figure}[H]
\begin{centering}
\vspace{-0.3cm}
\includegraphics[scale=0.35]{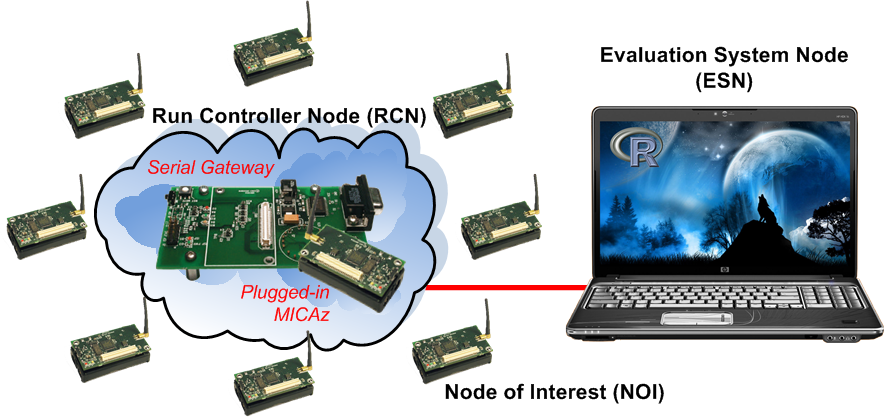}
\par\end{centering}

\caption{{\small \label{fig:Experimental-Testbed-Setup}Experimental testbed
setup and network topology}}
\vspace{-0.4cm}
\end{figure}

The experimental testbed setup includes several MICAz motes, a gateway/programming
board along with a Linux-based machine serving as \emph{Evaluation
System Node (ESN)}. One mote being attached to the gateway, henceforth
\emph{Run Controller Node (RCN)}, is in charge of test run coordination
and relaying of results to the ESN via its serial connection interface.
All other motes including the \emph{Node of Interest (NOI)} are endued
with experiment run-specific execution settings and act as parts of
a sample homogeneous WSN providing the RCN with feedback on gathered
results. To this end, the RCN polls the NOI every now and then between
experimental runs, collects measurement data and passes it to the
ESN (see Figure \ref{fig:Experimental-Testbed-Setup}). 

During experiment run series, the RCN triggers the sending process
of active motes by consecutively broadcasting control packets towards
the network. This emulates an event shower situation, i.e., a kind
of worst-case scenario for regional WSN operation, which implies concurrent
action of topographically close entities. In that, we optimize our
setup to isolate the actual cause for the phenomenon of packet loss
being pure medium access collisions of involved nodes. Therefore,
the broadcasts are guarded by 100 ms of inter-trigger lag to avoid
node event overlaps. Hence, observations are unbiased which we support
by a high number of samples helping to obtain fine-grained results
for univocal analysis. In addition, the RCN acts as the receiving
destination node in order to infer the PLR. For this purpose, the
RCN keeps track of the number of received packets originated by the
NOI. The measurement of the PSD, on its part, is conducted on the
NOI during the sending process triggered by RCN control packets. Its
value is included in the subsequent packet towards the RCN. The captured
PSD covers the time duration from the initiation of the packet delivery
process via the default TinyOS sending interface up to the point when
the event of successful modulation of the last packet bit is signaled
by lower-layer TinyOS components.

\begin{figure*}
\subfloat[\label{fig:PSD-BP_PS_20}Packet Size $P_{S}=20$ Byte]{\begin{raggedright}
\includegraphics[bb=40bp 0bp 1024bp 1024bp,scale=0.165]{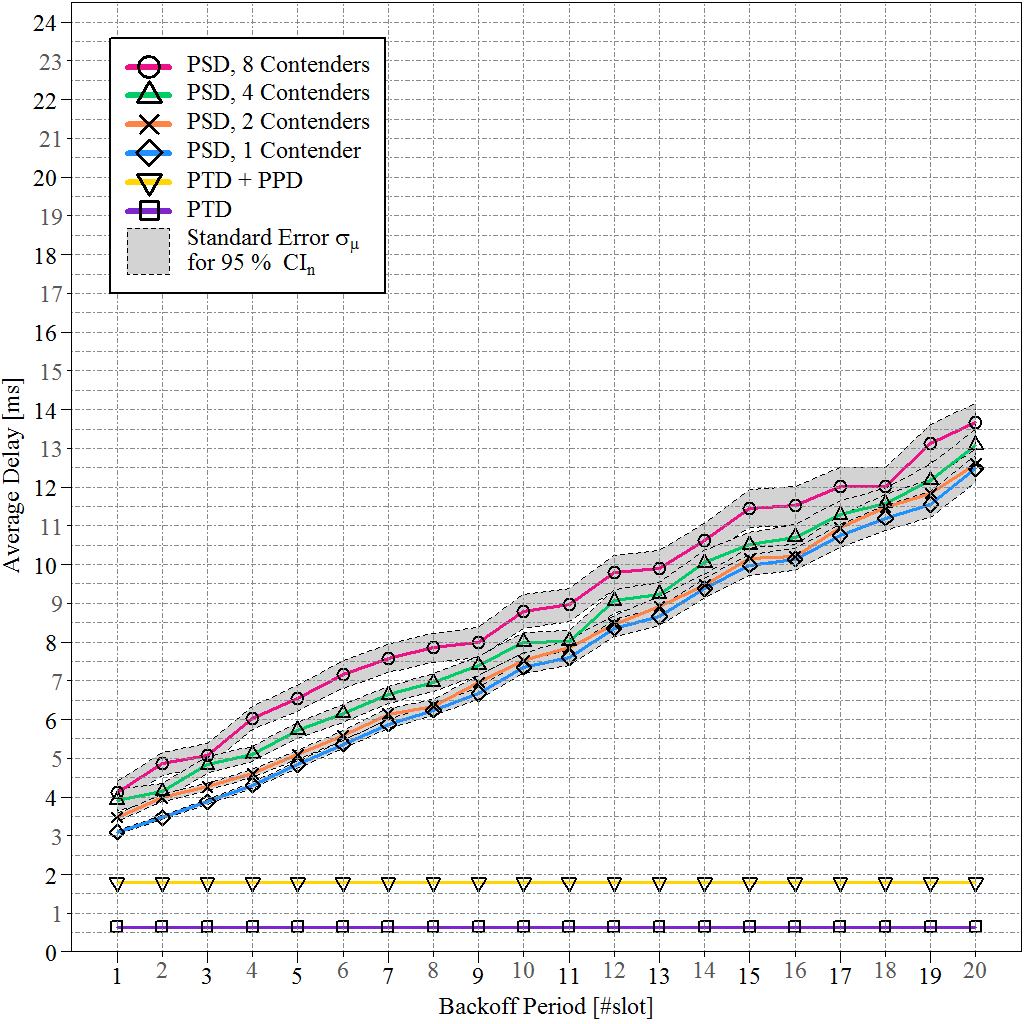}
\par\end{raggedright}

}\subfloat[\label{fig:PSD-BP_PS_70}Packet Size $P_{S}=70$ Byte]{\begin{raggedright}
\includegraphics[bb=40bp 0bp 1024bp 1024bp,scale=0.165]{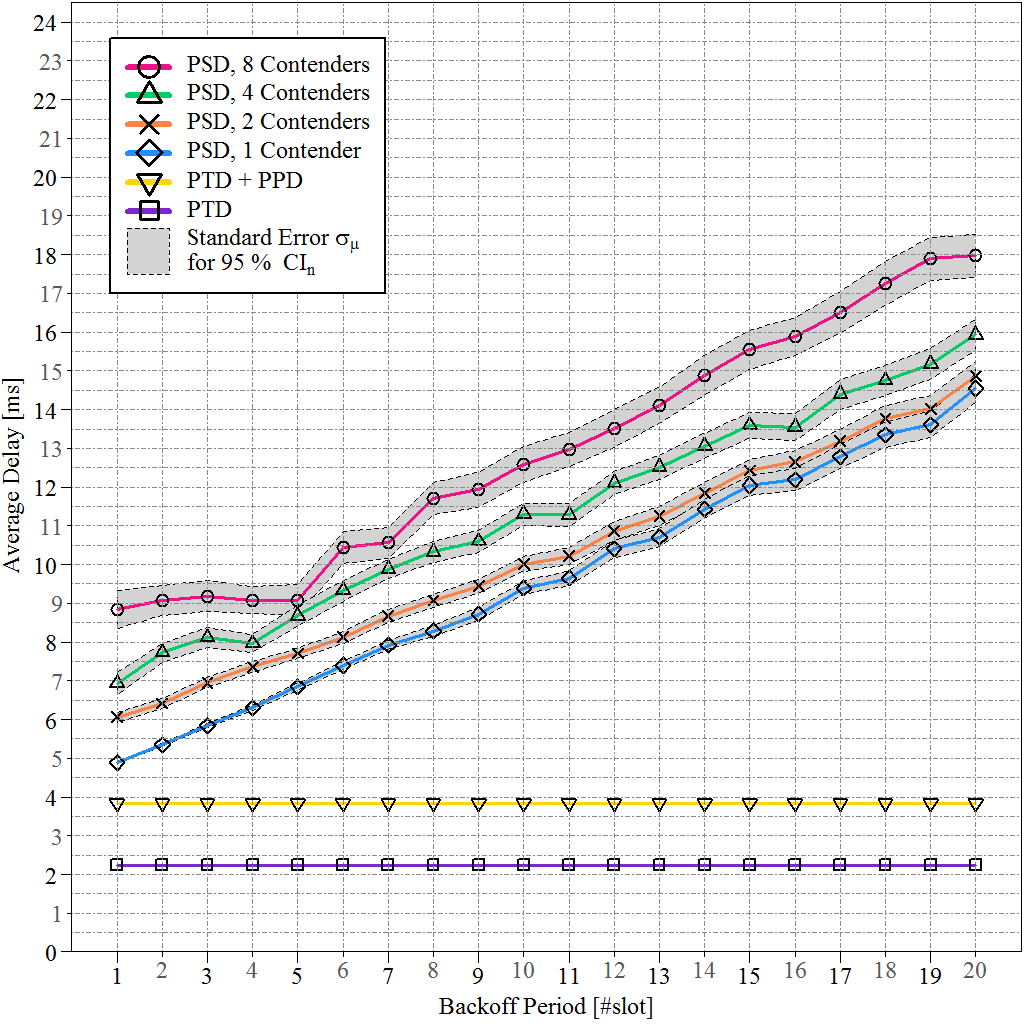}
\par\end{raggedright}

}\subfloat[\label{fig:PSD-BP_10_PS}Backoff Period $B_{P}=10$]{\begin{raggedright}
\includegraphics[bb=40bp 0bp 1024bp 1024bp,scale=0.165]{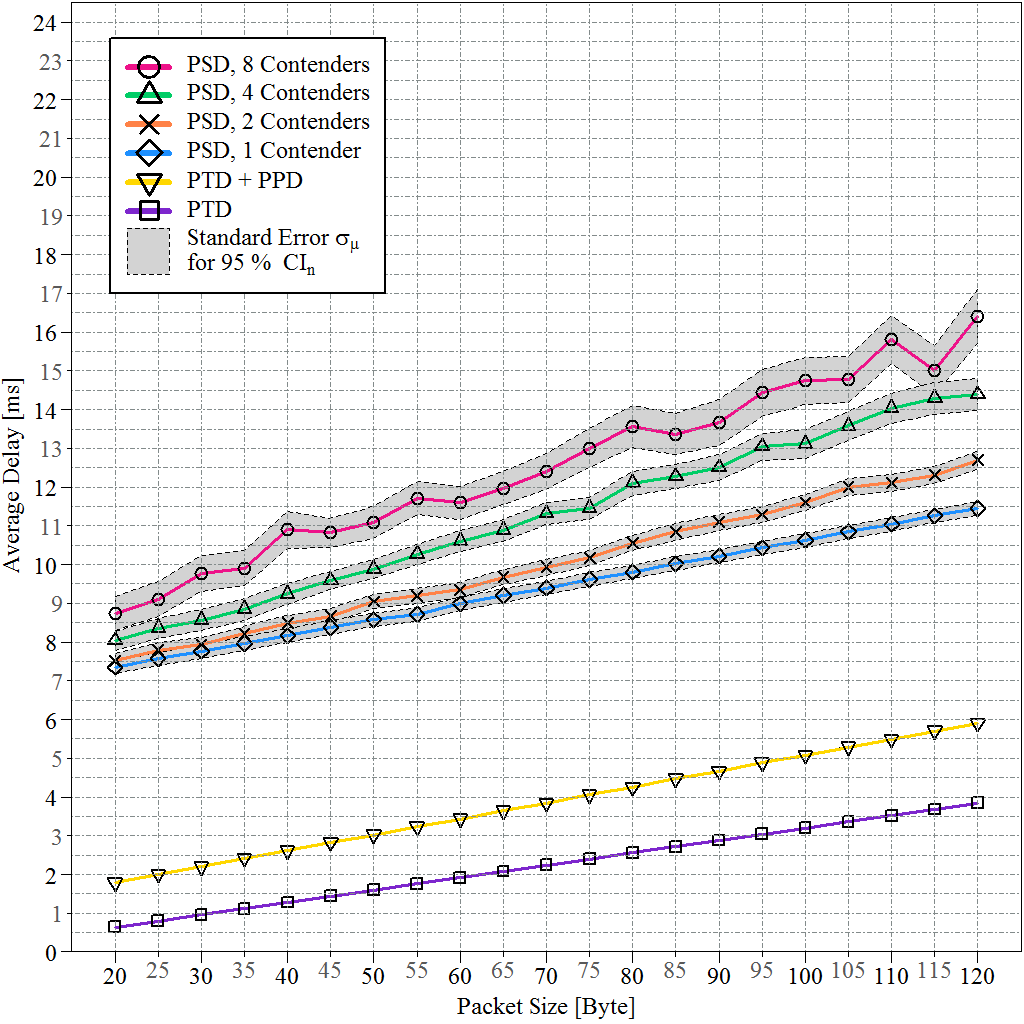}
\par\end{raggedright}

}

\caption{{\small \label{fig:Influence-on-PSD}Influence of backoff period,
packet size, and number of contenders on packet transmission, processing,
and sending delay}}
\vspace{-0.6cm}
\end{figure*}

Throughout all experimental runs, the three selected network parameters
are systematically permuted over the specified range of values (confer
Figure \ref{tab:General-and-experimental-configuration-settings}).
Therefore, the backoff period $B{}_{p}$ and packet size $P{}_{S}$
are programmatically adjusted whereas the number of contending nodes
$N{}_{C}$ is controlled by turning them on/off manually. In order
to obtain unambiguous information particularly on the number of lost
data packets with high granularity, we set the number of triggered
sending events to 1000 for each parameter constellation per experimental
run.

\subsection{Influence of B\emph{$_{P}$}, P$_{S}$ and N$_{C}$ on Average Delay}

As already indicated in Section II, the settings of the network parameters
is assumed to have a significant impact on the performance properties
under investigation. In order to proceed systematically, all influencing
factors are regarded separately. To this end, Figures \ref{fig:Influence-on-PSD}
and \ref{fig:Influence-on-PLR} include a subset of all experiment
results for representative parameter constellations. 

Figure \ref{fig:Influence-on-PSD} \subref{fig:PSD-BP_PS_20} reveals
the developing of mean delays against an increasing backoff period
also in view of four contender configurations while the packet length
is minimized to 20~Byte. As can be clearly seen, the PSD increases
gradually with all settings of the backoff period. Although it also
shows a virtual dependency on the number of neighbors, the confidence
intervals for 0.05 significance level overlap especially for higher
backoff period values, signifying its statistical indistinguishability
when it comes to the standard error of the computed mean from the
real population mean. This is due to the increasing standard deviation
of the samples lying between 776~\textmu{}s and 5729~\textmu{}s
for no neighbors and between 1324~\textmu{}s and 6044~\textmu{}s
for 7 neighbors, respectively, as far as backoff periods between 1
and 20 slots are concerned. This also emphasizes the high probabilistic
impact, albeit ascertainable as almost linear, of the backoff period
parameter on the overall delay. A more convincing statistical relevance
of the number of contenders becomes evident not before the packet
size is increased as exemplified for $P{}_{S}=70$~Byte in Figure
\ref{fig:Influence-on-PSD} \subref{fig:PSD-BP_PS_70}, where, additionally,
a slightly wider spreading of the confidence intervals and also a
definite elevation of the PSD can be observed. The latter trend, which
obviously suggest a sensitivity of the PSD to the packet length, is
captured in Figure \ref{fig:Influence-on-PSD} \subref{fig:PSD-BP_10_PS}.
In terms of the default backoff period setting of 10~slots and incrementation
of the packet length by 5~Byte over the available range of values,
the average PSD increases apparently in a linear manner as can be
concluded visually from the diagram. Again, a severe statistical relevance
of the number of contenders for the mean PSD, at least for the first
two proximate instances of the traversed value range, cannot be assumed
before a packet size of 50~Byte is reached. 

For the sake of completeness, it shall be noted that throughout all
experimental runs, we observed the devolution of the PPD (see Section
2.2) as well. It turned out that the PPD is exclusively dependent
on the packet size $P{}_{S}$ as expected. However, in order to shed
light on its portion of the overall time readings in isolation, we
conducted disjoint measurement runs freezing the random number to
$r=100$ for $B{}_{P}=10$ within the backoff process, i.e., observing
a constant MAD of 3520~\textmu{}s for an initial backoff value of
$z=31$ as applied on first transmission attempt in case of only one
sender. In this way, we numerically divided the PSD into its individual
components, i.e., PPD, MAD, PTD, the relationship of which becomes
evident in Figures \ref{fig:Influence-on-PSD} \subref{fig:PSD-BP_PS_20}-\subref{fig:PSD-BP_10_PS}.
Whereas the PTD can be assumed to be consistently dependent on the
transceiver transmission speed, the measurements of the PPD yield
a median standard deviation of about 11~\textmu{}s proving its constancy
in view of any parameter constellation. In summary, we can act on
the assumption of a linearly proportional interconnection between
the packet size $P{}_{S}$ and not only the PTD but also the PPD the
latter of which can be attributed to the preparation and in-system
transfer of packet data between the microcontroller main memory and
the radio transceiver transmission buffer \cite{2008-Bachorek-EnablingAuthenticTransmissions}.

\subsection{Influence of B\emph{$_{P}$}, P$_{S}$ and N$_{C}$ on Packet Loss}

\begin{figure*}
\subfloat[\label{fig:PLR-BP_PS_20}Packet Size $P_{S}=20$ Byte]{\begin{raggedright}
\includegraphics[bb=40bp 0bp 1024bp 1024bp,scale=0.165]{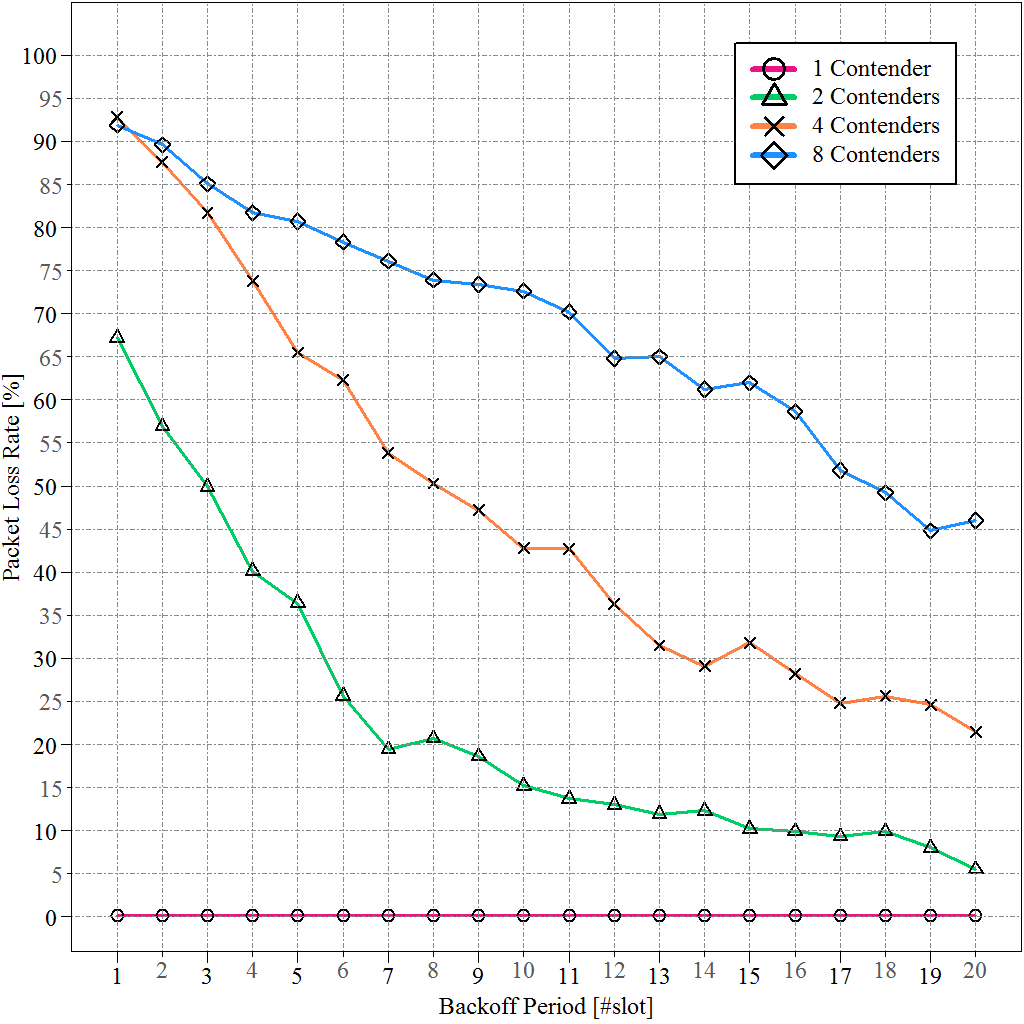}
\par\end{raggedright}

}\subfloat[\label{fig:PLR-BP_PS_70}Packet Size $P_{S}=70$ Byte]{\begin{raggedright}
\includegraphics[bb=40bp 0bp 1024bp 1024bp,scale=0.165]{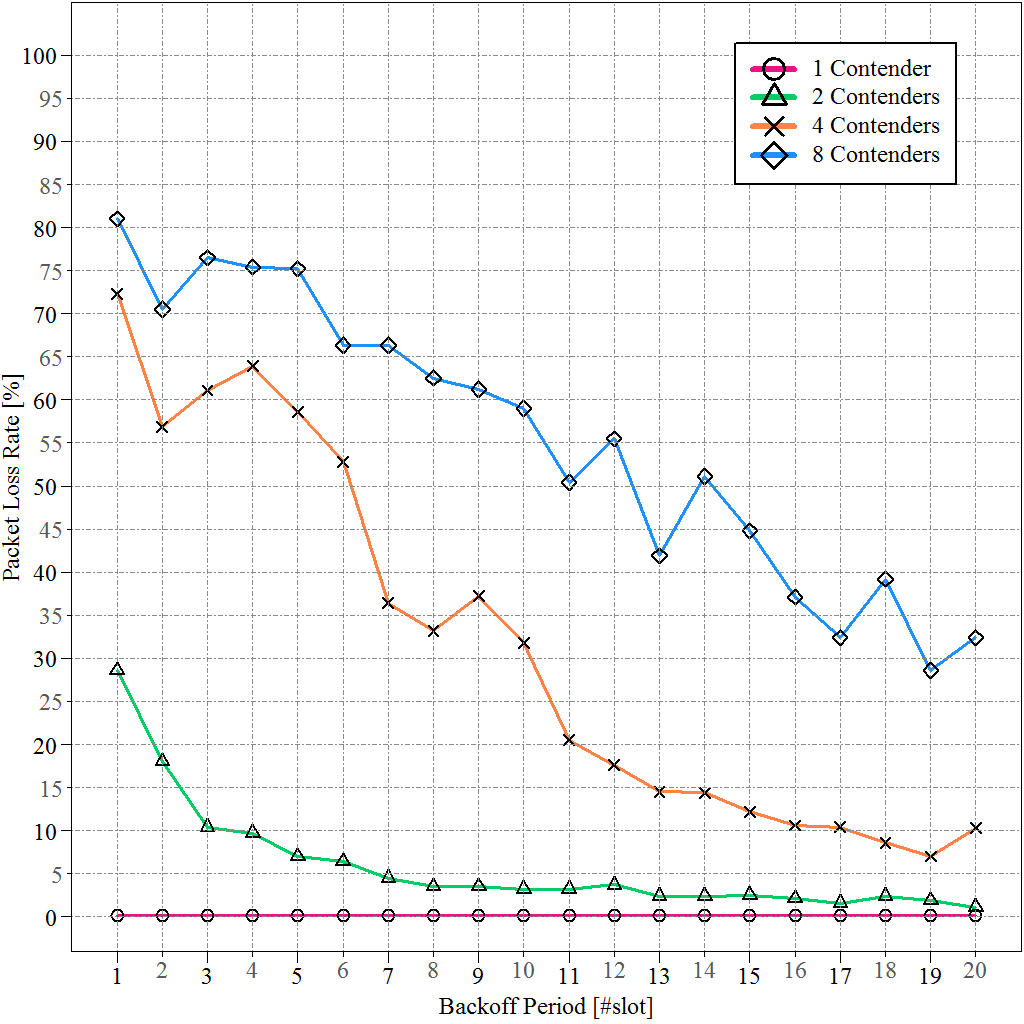}
\par\end{raggedright}

}\subfloat[\label{fig:PLR-BP_10_PS}Backoff Period $B_{P}=10$]{\begin{raggedright}
\includegraphics[bb=40bp 0bp 1024bp 1024bp,scale=0.165]{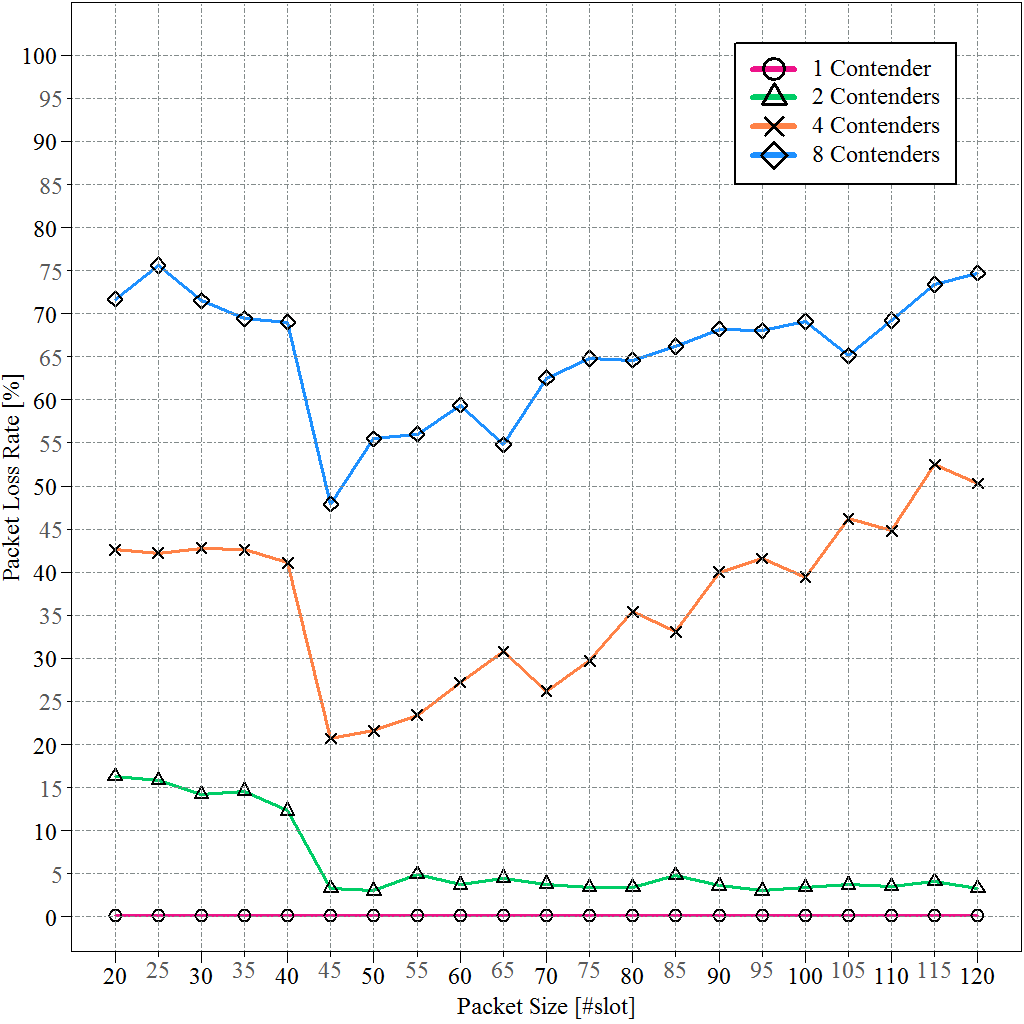}
\par\end{raggedright}

}

\caption{{\small \label{fig:Influence-on-PLR}Influence of backoff period,
packet size, and number of contenders on packet loss rate }}
\vspace{-0.6cm}
\end{figure*}

In contrast to the results on the PSD, the impact of the tested network
parameters exhibits a partly different behavior with respect to the
PLR. Intuitively, a wider range of the backoff delay interval implies
a lesser chance of packet collisions in case more than one sender
is actively probing for medium access. In effect, our field tests
show a proportionally increasing PLR for an increased number of channel
contenders $N{}_{C}$ as depicted by Figure \ref{fig:Influence-on-PLR}
which is persistently true for all parameter constellations. While
a single sending node witnesses a PLR of approx. 0.1~\% on the average
due to system internal inaccuracies, 2, 4 and 8 contending senders
can be subject to fluctuating PLRs of up to 92.8~\% clearly depending
on the backoff period and, at the first glance, also on the packet
size when comparing Figures \ref{fig:Influence-on-PLR} \subref{fig:PLR-BP_PS_20}
and \subref{fig:PLR-BP_PS_70}. However, the impact on PLR shows divergent
coherence in general when contrasting backoff period $B{}_{P}$ against
packet size $P{}_{S}$ for decoupled parameter iterations. While the
captured PLR obviously increases with a gradual decay of $B{}_{P}$
in Figure \ref{fig:Influence-on-PLR} \subref{fig:PLR-BP_PS_20} and
\subref{fig:PLR-BP_PS_70}, there is almost no indication for a simply
proportional relation between the PLR and $P{}_{S}$ as shown in Figure
\ref{fig:Influence-on-PLR} \subref{fig:PLR-BP_10_PS}. The only noticeable
observation remains the steady decline of the PLR from the highest
packet size down to 45~Byte which thereupon reverts to a sudden remarkable
increase followed by a nearly constant relaxation. This phenomenon
appears to be connected to the general presence of channel contenders
and might be accredited to an unbalanced random number generation
process with an unfavorable sequence of choices out of the value space.
Nonetheless, from a statistical point of view, $P{}_{S}$ appears
to play a marginal role with regard to the PLR when considered in
isolation. For this reason, we could limit our considerations to the
adjustment of the backoff period which proved to be a valuable driver
not only for the PSD but also the PLR. But, increasing PLR slope fluctuations
as observed for all backoff period iteration series in combination
with different packet sizes hypothesizes a potentially explicit correlation
of those two network parameters that requires further contemplation.
Other than that, the relatively small sample size for obtaining the
PLR might also add to the encountered discrepancy, not to mention
its implications on the precision of PLR predictions. In any case,
there is clear evidence for a negative correlation of the PSD and
the PLR which has to be taken into account when searching for the
most appropriate parameter constellation including $B{}_{P}$. Contrary,
the increase in number of contending nodes $N{}_{C}$ results not
only in an increased PLR but also in a higher mean PSD. In general,
a higher PLR also adds to the E2ED in that it usually implies increased
need for retransmissions and, thus, higher delays encountered at loss-intolerant
applications which rely on the end-to-end protocol argument. 

Summing up, the selected variety of presented experiment results reveals
a couple of valuable insights into the characteristics of a concrete
WSN node platform with regard to the I/EPSM approach. Based on these,
prospective network entities shall be enabled to derive performance
prediction models for the considered network properties next.

\section{Predicting Performance Properties in WSNs - A Lightweight Yet Powerful
Method}

As part of the analysis principle of the I/EPSM approach, the determination
of an appropriate performance model type along with its concrete instantiation
based on available measurement results constitute the fundamental
enabling procedure for meaningful estimation of performance property
values. A subsequent appraisal of enabled predictions against application
requirements, in turn, will allow for an appropriate adaptation of
selected network parameters for meeting imposed performance-related
demands in a terminal step of the I/EPSM task cycle, the treatment
of which, however, goes beyond the contents of this work. Also, any
other modeling methodologies for value prediction based on, e.g.,
Network Calculus or Artificial Neural Nets might come into question
as the foundation for the analysis principle further sophisticating
the I/EPSM approach in terms of its flexibility. Yet, we focus on
a lightweight yet powerful method as the most promising and demonstrably
viable approach, even from the intrinsic point of view, for now as
detailed in the following.

\subsection{Regression Modeling and Analysis}

One of the most common statistical methods for predicting random variables
used by analysts is based on regression modeling \cite{2006-Gelman-Dataanalysisusing}.
A variety of regression techniques such as (curve) linear, non-parametric,
mixed effects, a.o., exist to fit given observations of presumably
related data into a corresponding model \cite{1991-Jain-ArtofComputer}.
Usually, in order to identify and verify the right modeling technique
as part of the modeling task within the I/EPSM task cycle, statistical
tests are to be implemented on the corresponding network entities
in charge. However, in view of our experimental outcome from Section
III, simple and multiple linear regression emerge as most suitable
for finding the performance model instance $\Psi_{i}$ of choice that
relates the network properties from tuple $\left\langle \Upsilon_{t}\right\rangle =\left\langle PSD,PLR\right\rangle $,
each considered as predicted response $y$, to the nonstochastic network
parameters $x$ from set $\Phi=\left\{ B_{P},P_{S},N_{C}\right\} $,
that act as predicting explanatory variables according to Eq.\eqref{eq:Regression Formula}
with $m\in\left\{ 1,2,3\right\} $.\vspace{-0.3cm}

\begin{equation}
y_{_{\Upsilon_{t}}}=r_{0}+\left(\sum_{i=1}^{m}r_{i}\cdot x_{i}\right)+\epsilon\,\,\,,x_{i}\in\Phi\smallsetminus\bigcup_{k=1}^{i}x_{k-1}\label{eq:Regression Formula}
\end{equation}
where $x_{0}=\varnothing$, $t=\left\{ 0,1\right\} $, $r_{i}$ are
regression coefficients, and $\epsilon$ denotes the residual error
term.

In the course of singling out the best fitting model variant $\Psi_{i}$
for any of the 7 distinct parameter constellations per element of
$\Upsilon$, we consider our entire set of measurement results. Since
we accomplished 492 experimental observation runs in total covering
more than 29~\% of all possible combinations within the defined value
range bounds for all 3 network parameters while capturing up to 1000
samples per run, we can rely on 488 up to 490 degrees of freedom for
the statistical estimation of regression coefficients provided that
the model errors are independent and subject to a Gaussian distribution
with zero mean and constant standard deviation.

In the first place, we derive a simple linear regression model to
relate the properties in $\Upsilon$ to any individual network parameter
from $\Phi$ in a linear fashion. Subsequently, both network properties
are considered in view of all possible combinations of those parameters
by virtue of a multiple linear regression model each. In order to
find the mean predicted response with as little variability as possible,
we apply the least-squares criterion procedure minimizing the sum
of squared errors in view of the observed mean response $y_{j}$ as
per Eq. \ref{eq:SSE} for all n observations. The result summary is
tabulated in Figure \ref{fig:Statistical-significance-indicators-and-prediction-quality}
(left).

Expectedly, the results show that any of the network parameters has
a distinct influence on any of both properties to a certain extent
as can be told by the corresponding \emph{Coefficient of Determination
($\Omega$)}. In fact, \emph{$\Omega$$ $} discloses how much of
the variation of the response variable is explained by the regression
as the measure of choice when comparing the predictability relevance
of fitted regression values of different models \cite{1991-Jain-ArtofComputer}.
On this note, $\Psi_{1}$ based on just the backoff period appears
to have highest single explanatory rate of about 62.53~\% on the
PSD in contrast to the number of contenders $N{}_{C}$ that seems
to play just an inferior role with only 9~\% as already anticipated
in Section III. Nevertheless, its combination with the other two parameters
within model variant $\Psi_{7}$ is assumed to most precisely explain
the PSD on the average by 97.56~\%, i.e., it can predict its mean
value more reliably than all the other models do. On the other hand,
$\Psi_{3}$ turns out to have by far the highest explanatory value
of 65.82~\% regarding the PLR while the pure influence of the packet
size $P{}_{S}$ seems minuscule in accordance with our experimental
evaluation. Again, the combination of all three network parameters
in $\Psi_{7}$ results in the most optimistic model instance, albeit,
not significantly more than $\Psi_{5}$ which is based on just two
of these.\vspace{0cm}

\begin{equation}
min\left[\overset{n}{\underset{j=1}{\sum}}\left(y_{j}-y_{_{\Upsilon_{t}},j}\right)^{2}\mid\,\,\frac{1}{n}\underset{l=1}{\sum^{n}}\epsilon_{l}=0\right]\label{eq:SSE}
\end{equation}

\vspace{0cm}

\subsection{Performance Model Validation in Multi-Hop Scenarios}

From an analytical perspective, a best practice for choosing the right
model is to compare the statistical relevance of regression for all
constellations of available predictor variables. However, whereas
the implementation of deducing regression coefficients along with
\emph{$\Omega$} is rather straightforward, their further validation
implies more involved operations including, a.o., distribution generation,
normality proofing, and result transformation. Whereas some of these
may remain delegated to an external network entity subject to EPSM,
we focus on an approved subset of viable validation candidates for
further integration into IPSM-driven node implementations.

\begin{figure*}
\begin{raggedright}
{\footnotesize }%
\begin{minipage}[t]{1.35\columnwidth}%
\begin{flushleft}
{\footnotesize \vspace{0bp}
}%
\begin{tabular}{cccccc>{\centering}p{1.5cm}}
\toprule 
\textbf{\scriptsize Model} & \textbf{\scriptsize PSD | PLR} & \multicolumn{2}{c}{\textbf{\scriptsize Coefficient of Determination}} & \multicolumn{2}{>{\centering}p{2.2cm}}{\begin{spacing}{0.59999999999999998}
\textbf{\scriptsize \vspace{-0.4cm}
}\\
\textbf{\scriptsize Non-Zero Confidence Intervals\vspace{-1cm}
}\end{spacing}
} & \multirow{2}{1.5cm}{\begin{spacing}{0.59999999999999998}
\noindent \centering{}\textbf{\scriptsize \vspace{-0.6cm}
}\\
\textbf{\scriptsize Zero Parameter Correlation\vspace{-0.4cm}
}\end{spacing}
}\tabularnewline
\cmidrule{3-6} 
\textbf{\scriptsize $\Psi_{i}$} & \textbf{\scriptsize \textasciitilde{}} & \textbf{\scriptsize $\Omega_{PSD_{i}}$ {[}\%{]}} & \textbf{\scriptsize $\Omega_{PLR_{i}}$ {[}\%{]}} & \textbf{\scriptsize PSD} & \textbf{\scriptsize PLR} & \tabularnewline
\midrule
\midrule 
{\scriptsize 1} & {\scriptsize B$_{P}$} & \textbf{\scriptsize 62.53} & {\scriptsize 16.07} & \ding{51} & \ding{51} & -\tabularnewline
\midrule 
{\scriptsize 2} & {\scriptsize P$_{S}$} & {\scriptsize 26.03} & {\scriptsize 0.41} & \ding{51} & \ding{55} & -\tabularnewline
\midrule 
{\scriptsize 3} & {\scriptsize N$_{C}$} & {\scriptsize 9.00} & \textbf{\scriptsize 65.82} & \ding{51} & \ding{51} & -\tabularnewline
\midrule 
{\scriptsize 4} & {\scriptsize B$_{P}$ + P$_{S}$} & \textbf{\scriptsize 88.56} & {\scriptsize 16.48} & \ding{51}\ding{51} & \ding{51}\ding{55} & \ding{51}\tabularnewline
\midrule 
{\scriptsize 5} & {\scriptsize B$_{P}$ + N$_{C}$} & {\scriptsize 71.53} & \textbf{\scriptsize 81.88} & \ding{51}\ding{51} & \ding{51}\ding{51} & \ding{51}\tabularnewline
\midrule 
{\scriptsize 6} & {\scriptsize P$_{S}$ + N$_{C}$} & {\scriptsize 35.03} & {\scriptsize 66.22} & \ding{51}\ding{51} & \ding{51}\ding{51} & \ding{51}\tabularnewline
\midrule 
{\scriptsize 7} & {\scriptsize B$_{P}$ + P$_{S}$ + N$_{C}$} & \textbf{\scriptsize 97.56} & \textbf{\scriptsize 82.29} & \ding{51}\ding{51}\ding{51} & \ding{51}\ding{51}\ding{51} & \ding{51}\tabularnewline
\bottomrule
\end{tabular}
\par\end{flushleft}%
\end{minipage}%
\begin{minipage}[t]{1\columnwidth}%
\begin{flushleft}
{\footnotesize \vspace{0bp}
}\includegraphics[bb=0bp 0bp 1024bp 1024bp,scale=0.165]{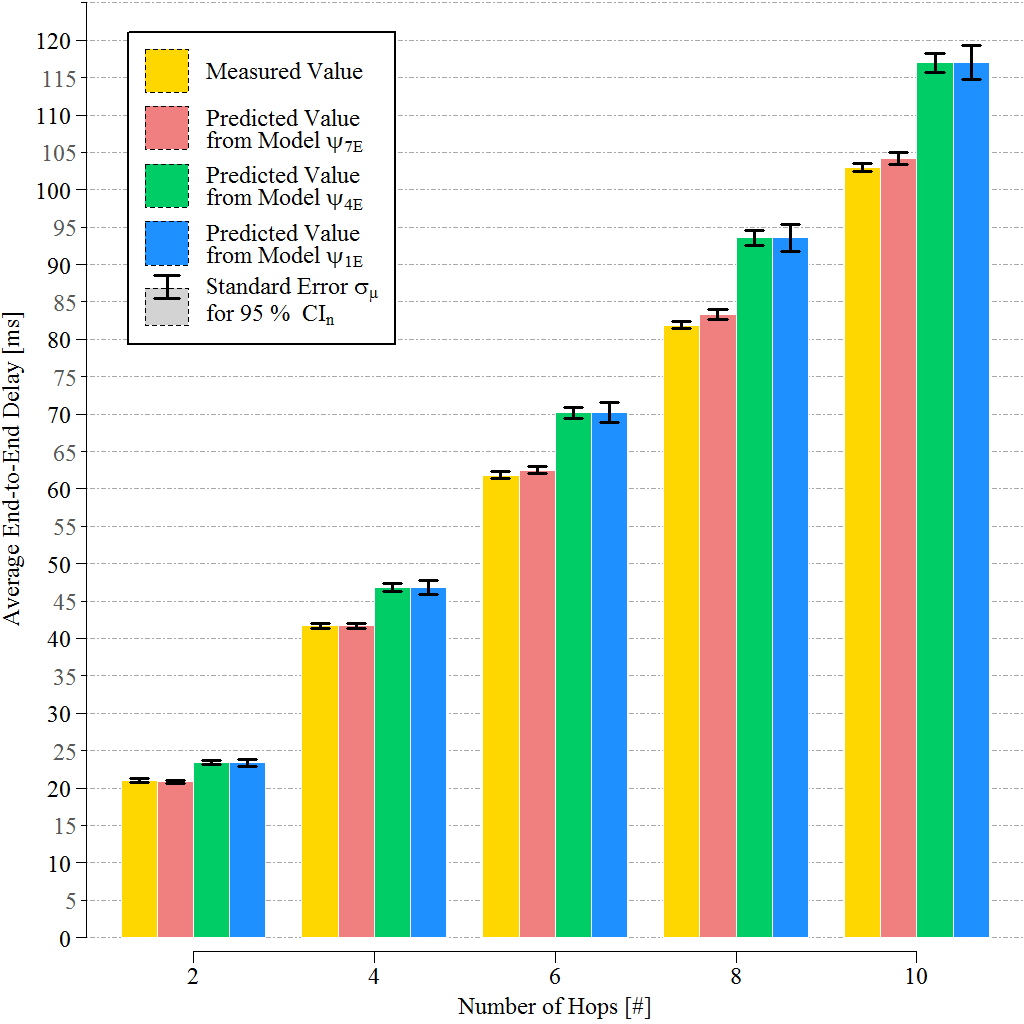}
\par\end{flushleft}%
\end{minipage}
\par\end{raggedright}

\caption{{\small \label{fig:Statistical-significance-indicators-and-prediction-quality}Statistical
significance indicators of linear regression models for PSD and PLR
prediction at 95~\% confidence level (left) and comparison of model
prediction quality for E2ED in multi-hop scenarios under defaults
B$_{P}=10$, P$_{S}=70$~Byte, and N$_{C}=1$ (right)}}
\vspace{-0.6cm}
\end{figure*}

In this context, examining the \emph{Confidence Intervals (CIs)} for
the regression coefficients of a model offers valuable clues to the
variability of these estimates. Concurrently, testing CIs also reveals
if the regression explains a significant part of the variation of
the response variable. In case the CI does not include zero, the coefficient
of the variable is said to be non-zero and thus the regression is
statistically significant and vice versa. Whereas this holds true
for even a 0.05 significance level for any single predictor variable
tested against PSD as in isolation so in combination, this assumption
cannot be generally anticipated for PLR estimation, particularly in
case of multiple response drivers as present within model option $\Psi_{4}$.
This is due to potential multicollinearity effects of the predictor
variables that counterintuitively might reduce the statistical accuracy
of the applied regression method. However, the sighted incongruity
regarding the packet size parameter merely reflects its irrelevance
in view of PLR prediction as already foreshadowed in the previous
section. Apart from that, all network parameters pass the corresponding
zero correlation test and clearly exhibit an additive cohesion throughout
all model alternatives, both of which indicate their uncorrelated
pertinence for response prediction (confer Figure \ref{fig:Statistical-significance-indicators-and-prediction-quality}
(left)). 

In the course of a proof of concept for the analysis principle wihtin
the IPSM inspection phase, we also implemented the linear regression
method on a real WSN mote. The results confirm that the alleged resource
limitations do not hamper the feasibility and precision of the very
modeling and prediction technique at all. However, the implementation
of a dynamic storage management necessary to handle the vast amount
of sample data effectively within the pretty slow flash storage, offering
a mean access time of about $4\, ms$ per observation, challenged
the hardware in terms of 25\% leftover main memory during execution.
A moving average function implemented to consider only the last captured
value into a continuously updated mean PSD for every network parameter
setting constellation from$\Phi$, turned out to be significantly
close to the mean over the entire measurement set. In this manner,
just a subset of values of $\left(\left\Vert \Phi\right\Vert +2\right)\cdot4\, Byte$
for each compound observation with \textmu{}s precision for PSD was
stored to conduct the sequence of calculations needed for regression
model instantiation\cite{1991-Jain-ArtofComputer}.

Since further considerations might imply more complex and partly divergent
decisions depending on the used modeling technique that go beyond
simple numerical operations as state before, we dedicate our regard
to a pragmatic comparison of enhanced prediction results to measured
E2ED values for some basic multi-hop WSN scenarios as follows.

Based on the original experiment setup, we organized hop-variant network
topologies of tandem-like shape and measured the E2ED (where PLR would
work analogously) between the source and the destination node for
default parameter settings as per Figure \ref{fig:Statistical-significance-indicators-and-prediction-quality}
(right). Apart from the PSD as estimated by a subset of most promising
model instances selected based on \emph{$\Omega$}, we amplify our
models by including the PPD, which is assumed to additionally incur
at every receiving node $r$, and by postulating an additive coherence
between the E2ED and the number of hops $h$ according to\vspace{0cm}
 
\begin{equation}
y_{_{\Upsilon_{1},E}}=h\cdot\left(y_{_{\Upsilon_{1}}}+\frac{1}{2}PPD_{r}\right)\label{eq:Enhanced model forumula}
\end{equation}
\vspace{0cm}

Indeed, yield results show an almost perfect forecast of the E2ED
for up to 10 hops using enhanced model $\Psi_{7E}$ as indicated by
overlapping CIs for 95~\% confidence level. Other model candidates,
in turn, seem to mismatch the E2ED to a large extent. Interestingly,
the prediction based on $\Psi_{1E}$ is equal to that based on $\Psi_{4E}$,
albeit, with a lower certainty, even though P$_{S}$ is unconsidered
in $\Psi_{1E}$ and \emph{$\Omega$$ $}$_{PSD_{1}}<\Omega_{PSD_{4}}$\emph{$ $}.
This example points out the importance of diversity in consulting
statistical measures when comparing prediction quality. Nevertheless,
a general trend towards overestimation of the E2ED by any model can
be observed that seems to sums up with the increase in number of hops.
This either suggests a missing influence factor or just a misconception
of the considered ones. In fact, in our case, the PPD as part of the
sending process verifiably differs from the PPD during reception by
about the half as prior node-local measurements revealed. Again, this
result illustrates that a performance model can only be as good as
the knowledge about the integral parts its prediction is based upon.
That is why, prospective self-managed network entities are meant to
collect or conclude as many details about their constituent parts
as possible adhering to the I/EPSM control principle.

Summing up, whereas model type selection and definite model validation
remain out of scope for resource-constrained devices for now, we verified
the investigated modeling technique to be what a WSN node is able
to apply for meaningful performance model derivation in virtue of
its given capabilities. This is in line with our original aim to keep
computational costs and complexity as rational as possible rendering
the methodology what is considered lightweight and eligible for the
I/EPSM concept.

\section{Conclusion and Future Work}

This work dealt with an introductory investigation of a measurement-based
approach to self-management of performance-related network properties
based upon a common methodology for stochastic value prediction. We
have conducted extensive measurements to determine the influence of
three selected network parameters on two fundamental network properties.
A concise overview of applied technology, experimental design, and
theoretical background introduced the basis for our drawn conclusions.
Our main results confirmed previous assumptions of a linear relationship
between packet sending delay and backoff period, packet size, as well
as number of contending nodes for diverse networking scenarios. Further
findings on packet loss rate and its relation to the parameters in
question have been obtained, e.g., proneness to fluctuations for certain
influence factor variations. Furthermore, we derived several regression
models based on empirical data enabling property forecast for arbitrary
parametrical settings even on a resource-scarce WSN mote. A subsequent
model validation followed by a practical verification of yield speculations
on performance behavior in basic multi-hop networks finally revealed
the high precision of an augmented performance model in view of end-to-end
delay prediction. All in all, we have demonstrated viable methods
for the aspired measurement and modeling tasks based on principles
as part of our comprehensive I/EPSM approach that bears good prospects
due to its flexibility in further integration of analysis methods
along with load-balanced collaboration techniques in future concretions.
In this context, we aim at considering lightweight procedures with
as little demands on computational resources as possible by using
network inherent information that can be drawn from regular network
node operation and communication behavior. 

As part of future work, we intend to explore further WSN peculiarities
as influencing factors, e.g., duty cycling and data-aggregation, and
also test the quality of prediction for network properties other than
the ones considered herein. Also, various other networking scenarios
shall be subject to complementary future investigations. In order
to cover more involved network topologies and parameter interactions,
simulations constitute the basis for obtaining comparative results
for larger-scale networks. Finally, in view of our long-term research
endeavors, the implementation of sophisticated routines for model
validation purposes along with the integration of other property prediction
techniques on a unified infrastructure for network autonomy are also
up to future work. 

\bibliographystyle{ieeetr}
\bibliography{arXiv-References}

\begin{thebibliography}{10}

\bibitem{2011-Buchmayr-surveysituationaware}
M.~Buchmayr and W.~Kurschl, ``{A Survey on Situation-aware Ambient Intelligence
  Systems},'' {\em Springer Journal of Ambient Intelligence and Humanized
  Computing}, vol.~2, pp.~175--183, June 2011.

\bibitem{2006-Benini-Wirelesssensornetworks:}
L.~Benini, E.~Farella, and C.~Guiducci, ``{Wireless Sensor Networks: Enabling
  Technology for Ambient Intelligence},'' {\em Elsevier Microelectronics
  Journal}, vol.~37, pp.~1639--1649, September 2006.

\bibitem{2002-Akyildiz-WirelessSensorNetworks}
I.~F. Akyildiz, W.~Su, Y.~Sankarasubramaniam, and E.~Cayirci, ``{Wireless
  Sensor Networks - A Survey},'' {\em Elsevier Computer Networks}, vol.~38,
  pp.~393--422, March 2002.

\bibitem{2007-Akyildiz-surveywirelessmultimedia}
I.~F. Akyildiz, T.~Melodia, and K.~R. Chowdhury, ``{A Survey on Wireless
  Multimedia Sensor Networks},'' {\em Elsevier Computer Networks}, vol.~51,
  pp.~921--960, January 2007.

\bibitem{2012-Movahedi-SurveyAutonomicNetwork}
Z.~Movahedi, M.~Ayari, R.~Langar, and G.~Pujolle, ``{A Survey of Autonomic
  Network Architectures and Evaluation Criteria},'' {\em IEEE Communications
  Surveys and Tutorials}, vol.~14, pp.~464--490, June 2012.

\bibitem{2011-KostasTsagkaris-Autonomicsinwireless}
K.~Tsagkaris, P.~Vlacheas, G.~Athanasiou, V.~Stavroulaki, S.~Filin, H.~Harada,
  J.~Gebert, and M.~Mueck, ``{Autonomics in Wireless Network Management:
  Advances in Standards and Further Challenges},'' {\em IEEE Network}, vol.~25,
  pp.~41--49, November 2011.

\bibitem{2003-Ganek-dawningautonomiccomputing}
A.~G. Ganek and T.~A. Corbi, ``{The Dawning of the Autonomic Computing Era},''
  {\em IBM Systems Journal}, vol.~42, pp.~5--18, January 2003.

\bibitem{2005-Kephart-Researchchallengesautonomic}
J.~O. Kephart, ``{Research Challenges of Autonomic Computing},'' in {\em
  Proceedings of 27th International Conference on Software Engineering (ICSE
  2005)}, pp.~15--22, May 2005.

\bibitem{2009-Samaan-TowardsAutonomicNetwork}
N.~Samaan and A.~Karmouch, ``{Towards Autonomic Network Management: an Analysis
  of Current and Future Research Directions},'' {\em Communications Surveys \&
  Tutorials, IEEE}, vol.~11, pp.~22--36, August 2009.

\bibitem{2007-BrendanJennings-Towardsautonomicmanagement}
B.~Jennings, S.~v.d. Meer, S.~Balasubramaniam, D.~Botvich, M.~O. Foghlu,
  W.~Donnelly, and J.~Strassner, ``{Towards Autonomic Management of
  Communications Networks},'' {\em IEEE Communications Magazine}, vol.~45,
  pp.~112--121, October 2007.

\bibitem{2013-Volosencu-Efficiencyimprovementin}
C.~Volosencu and D.-I. Curiac, ``{Efficiency Improvement in Multi-sensor
  Wireless Network Based Estimation Algorithms for Distributed Parameter
  Systems},'' {\em Journal on Advances in Signal Processing}, January 2013.

\bibitem{2005-Lim-Energyefficientself}
J.-J. Lim and K.~G. Shin, ``{Energy-efficient Self-adapting Online Linear
  Forecasting for Wireless Sensor Network Applications},'' in {\em Proceedings
  of IEEE International Conference on Mobile Adhoc and Sensor Systems
  Conference}, November 2005.

\bibitem{2007-Schmitt-ComprehensiveWorstCase}
J.~B. Schmitt, F.~A. Zdarsky, and L.~Thiele, ``{A Comprehensive Worst-Case
  Calculus for Wireless Sensor Networks with In-Network Processing},'' in {\em
  Proceedings of 28th IEEE Real-Time Systems Symposium (RTSS)}, December 2007.

\bibitem{2011-RalfLuebben-foundationstochasticbandwidth}
R.~Lubben, M.~Fidler, and J.~Liebeherr, ``{A Foundation for Stochastic
  Bandwidth Estimation of Networks with Random Service},'' in {\em Proceedings
  of INFOCOM, 2011}, pp.~1817--1825, April 2011.

\bibitem{2011-Peterson-Computernetworks:systems}
L.~Peterson and B.~Davie, {\em {Computer Networks: A Systems Approach}}.
\newblock Morgan Kaufmann, 5th~ed., March 2011.

\bibitem{2006-Kurose-ComputerNetworks:Top}
J.~Kurose and K.~Ross, {\em {Computer Networks: A Top Down Approach Featuring
  the Internet}}.
\newblock Pearson Addison Wesley, 5th~ed., March 2009.

\bibitem{2007-Karl-Protocolsandarchitectures}
H.~Karl and A.~Willig, {\em {Protocols and Architectures for Wireless Sensor
  Networks}}.
\newblock Wiley-Interscience, 1st~ed., September 2007.

\bibitem{2007-TexasInstruments-ChipconCC2420-}
{Texas Instruments}, {\em {"Chipcon CC2420 - 2.4 GHz IEEE 802.15.4 /
  ZigBee-ready RF Transceiver"}}.
\newblock Texas Instruments Inc., March 2007.

\bibitem{2003-Schiller-Mobilecommunications}
J.~Schiller, {\em {Mobile Communications}}.
\newblock Addison Wesley, 2nd~ed., August 2003.

\bibitem{2003-IEEE-802.15.4-Standard}
IEEE, ``{802.15.4 - Standard for Local and Metropolitan Area Networks - Part
  15.4: Specifications for Low-Rate WPANs},'' October 2003.

\bibitem{2006-Crossbow-MoteProcessorRadio}
{Crossbow Technology Inc.}, {\em {"Mote Processor Radio (MPR) Platforms and
  Mote Interface Boards (MIB), Rev. B"}}.
\newblock Manual, June 2006.

\bibitem{2003-Gay-nesCLanguage-Holistic}
D.~Gay, P.~Levis, J.~R. von Behren, M.~Welsh, E.~A. Brewer, and D.~E. Culler,
  ``{The nesC Language: A Holistic Approach to Networked Embedded Systems},''
  in {\em Proceedings of the ACM Conference on Programming Language Design and
  Implementation (PLDI '03)}, ACM, June 2003.

\bibitem{2004-Polastre-Versatilelowpower}
J.~Polastre, J.~L. Hill, and D.~E. Culler, ``{Versatile Low Power Media Access
  for Wireless Sensor Networks},'' in {\em Proceedings of the 2nd IEEE
  International Conference on Embedded Networked Sensor Systems (SenSys)},
  pp.~95--107, November 2004.

\bibitem{2009-Levis-TinyOSprogramming}
P.~Levis and D.~Gay, {\em {TinyOS Programming}}, vol.~1.
\newblock Cambridge University Press, March 2009.

\bibitem{2007-Alliance-ZigBeeandWireless}
{ZigBee Alliance}, ``{ZigBee and Wireless Radio Frequency Coexistence},'' June
  2007.

\bibitem{2008-Bachorek-EnablingAuthenticTransmissions}
A.~Bachorek, I.~Martinovic, and J.~B. Schmitt, ``{Enabling Authentic
  Transmissions in WSNs - Turning Jamming against the Attacker},'' in {\em IEEE
  ICNP 4th Workshop on Secure Network Protocols (NPSec)}, October 2008.

\bibitem{2006-Gelman-Dataanalysisusing}
A.~Gelman and J.~Hill, {\em {Data Analysis Using Regression and
  Multilevel/hierarchical Models}}.
\newblock Cambridge University Press, August 2006.

\bibitem{1991-Jain-ArtofComputer}
R.~K. Jain, {\em {The Art of Computer Systems Performance Analysis: Techniques
  for Experimental Design, Measurement, Situation-awaremulation, and
  Modeling}}.
\newblock New York, NY (USA): John Wiley \& Sons, Inc., 1st~ed., April 1991.

\end{thebibliography}

\end{document}